# Quantum diffusion for a quantum particle with a correlated Gaussian noise

Yun Jeong Kang [a], Sung Kyu Seo [b,] and Kyungsik Kim [c,*]

[a] *School of Liberal Studies, Wonkwnag University, 54538, Republic of Korea*

[b] *Haena Ltd., Seogwipo-si, Jejudo 63568, Republic of Korea*

[c] *Department of Physics, Pukyong National University, Busan 48513, Republic of Korea*

We investigate the diffusive behavior of a quantum particle driven by a correlated Gaussian noise. We derive the analytical solution of the joint probability density function and obtain explicit expressions for the mean square momentum and the mean square displacement. In the short-time regime, the mean square momentum exhibits a super-ballistic scaling proportional to $t^2$ in the presence of a correlated Gaussian noise, whereas it scales as $t$ in the long-time limit. Furthermore, the mean square displacement grows as $t^4$ in the short-time regime and crosses over to a $t^3$ scaling under an uncorrelated (random) noise in the long-time limit. These results reveal the significant role of temporal noise correlations in modifying the quantum diffusion dynamics.

---

* Corresponding authors. E-mail: kskim@pknu.ac.kr

## 1. Introduction

The quantum mechanical motion of a particle in a dynamically disordered medium has attracted considerable attention in various continuum models [1–3]. In particular, Jayannavar and Kumar [4] reported a striking nondiffusive behavior characterized by a mean squared displacement $\langle x^2(t)\rangle \sim t^4 (t \to \infty)$, in contrast to the diffusive behavior $\langle x^2(t)\rangle \sim t$ generally expected from earlier studies of electrons in tight-binding lattices [5,6]. Their result suggested that dynamical disorder can fundamentally alter long-time transport properties. However, the physical origin of this anomalous scaling [7-9] was discussed only at a semi-phenomenological level.

From a classical perspective, consider the stochastic equation $m\frac{dv}{dt} = \xi(t)$, with the white noise satisfying $\langle \xi(t)\xi(t')\rangle = 2k_B T\gamma\delta(t-t')$. One readily obtains $\langle x^2(t)\rangle \propto t^3$, indicating nondiffusive superballistic growth. In this case, the particle continuously absorbs energy from the fluctuating force and accelerates indefinitely. This classical analogy suggests that anomalous quantum transport may originate from the interplay between noise correlations and inertial dynamics.

Despite extensive discussions of nondiffusive quantum transport, exact analytical treatments of quantum diffusion driven by a correlated Gaussian noise remain limited. In this Letter, we derive analytical solutions of the evolution equation for the joint characteristic function and obtain explicit expressions for the mean square displacement and momentum. We show that the short- and long-time behaviors exhibit distinct power-law scalings, clarifying the dynamical origin of superballistic growth in both coordinate and momentum space.

## 2. Quantum diffusion for a quantum particle with a correlated Gaussian noise

By converting the Schrödinger equation into momentum space representation, we express in the following integral equation form:

$$i\hbar\frac{\partial}{\partial t}\Psi(p,t) = \frac{p^2}{2m}\Psi(p,t) + \int_{-\infty}^{\infty} V(p-p',t)\Psi(p',t)\frac{dp'}{2\pi\hbar}. \quad (1)$$

To express this wave function as a function of momentum instead of position from Eq. (1), using the Fourier transform, we typically obtain Eq. (1), which is an integral equation. In the momentum space, the Schrödinger equation for a quantum particle corelated Gaussian force is given by

$$i\hbar\frac{\partial}{\partial t}\Psi(p,t) = \frac{p^2}{2m}\Psi(p,t) + \int_{-\infty}^{\infty} G(p-p'',t)\Psi(p'',t)\frac{dp''}{2\pi\hbar}. \quad (2)$$

We assume to be correlated that a quantum particle moving in a corelated Gaussian force $G(x,t)$ with

$$< G(p,t)G(p',t') >= \frac{G_0^2}{2\tau} g(p-p')\, exp(-|\, t-t'|/\tau) \tag{3}$$

Introducing $< F(p,p',t) >$ as $F(p,p',t) = \Psi(p,t)\Psi^*(p',t)$, we get the equation of motion for momentum space as

$$\frac{\partial}{\partial t} < F(p,p',t) >= \left(\frac{p^2}{2i\hbar m} - \frac{p'^2}{2i\hbar m}\right) < F(p,p',t) > + \frac{\beta}{2\pi\hbar} b(t)$$

$$\int_{-\infty}^{\infty} dp'' \int_{-\infty}^{\infty} dp''' [g(p-p''+p''')g(p'-p''') - g(0)] < F(p'',p''',t)>. \tag{4}$$

where $\beta = \frac{G_0^2}{2\hbar^2}$ and $b(t) = 1 - \exp(-t/\tau)$ for a quantum particle with a correlated Gaussian noise. When we put $p_{x_1} = p + p'$ and $p_{x_2} = p - p'$, we derive as

$$\frac{\partial}{\partial t} F(p_{x_1}, p_{x_2}, t) = \frac{1}{2i\hbar m} p_{x_1} p_{x_2} F(p_{x_1}, p_{x_2}, t) + \frac{\beta}{2\pi\hbar} b(t) \int_{-\infty}^{\infty} p'' \int_{-\infty}^{\infty} dp''' [g(p_{x_2} - p' + p''') - g(0)] F(p'', p''', t). \tag{5}$$

Taking the Fourier transforms in Eq. (5), the rearranging equation [10-12] is given by

$$\frac{\partial}{\partial t} F(\xi_1, \xi_2, t) = -\frac{2i\hbar}{m} \frac{\partial^2}{\partial \xi_1 \partial \xi_2} F(\xi_1, \xi_2, t) + \beta b(t) [g(\xi_2) - g(0)] F(\xi_1, \xi_2, t). \tag{6}$$

where the Fourier transform $F(\xi_1, \xi_2, t)$ of a wave function in momentum space is defined as follows:

$$F(\xi_1, \xi_2, t) = \frac{1}{2\pi\hbar} \int_{-\infty}^{\infty} dp_{x_1} \int_{-\infty}^{\infty} dp_{x_2} F(p_{x_1}, p_{x_2}, t) e^{-ip_{x_1}\xi_1/\hbar - ip_{x_2}\xi_2/\hbar}. \tag{7}$$

Now, we derive the solutions of the probability densities $F(\xi_1, t)$ and $F(\xi_2, t)$ in the short-time regime $t \ll \tau$. To find the special solutions for $\xi_1$ and $\xi_2$ by the variable separation from Eq. (6), the two equations are written as

$$\frac{\partial}{\partial t} F(\xi_1, t) = -D_{\xi_2} \frac{\partial}{\partial \xi_1} F(\xi_1, t) + \frac{\beta b(t)}{2} [g(\xi_2) - g(0)] F(\xi_1, t) + AF(\xi_1, t), \tag{8}$$

$$\frac{\partial}{\partial t} F(\xi_2, t) = -D_{\xi_1} \frac{\partial}{\partial \xi_2} F(\xi_1, t) + \frac{\beta b(t)}{2} [g(\xi_2) - g(0)] F(\xi_2, t) - AF(\xi_2, t), \tag{9}$$

where $A$ denotes the separation constant, $D_{\xi_1} \equiv \frac{i\hbar}{m} \frac{\partial}{\partial \xi_1}$, and $D_{\xi_2} \equiv \frac{i\hbar}{m} \frac{\partial}{\partial \xi_2}$. The function $g(\xi_2)$ denotes $g(\xi_2) = [2\pi\sigma_0^2]^{-1/2} \exp(-\xi^2 / 2\sigma_0^2)$.

Taking $\frac{\partial}{\partial t} F^{st}(\xi_1, t) = 0$ in the steady state, we obtain $F^{st}(\xi_1, t)$ as

$$F^{st}(\xi_1, t) = \exp\,[-\frac{\beta b(t)}{2D_{\xi_2}} [g(0) - g(\xi_2)] \xi_1 - \frac{2A}{D_{\xi_2}} \xi_1]. \tag{10}$$

To find the solution of the probability density for $\xi_1$ from $F(\xi_1, t) \equiv F^{st}(\xi_1, t) F^{st}(\xi_1, t)$, we include terms up to order $1/\tau_{th}^2$ and write

$$F(\xi_1, t) = H(\xi_1, t) \exp\,[-\frac{\beta b(t)}{2D_{\xi_2}} [g(0) - g(\xi_2)] \xi_1 - \frac{2A}{D_{\xi_2}} \xi_1], \tag{11}$$

$$H(\xi_1, t) = I(\xi_1, t) \exp\,[+\frac{\beta b'(t)}{2D_{\xi_2}^2} [g(0) - g(\xi_2)] \frac{\xi_1^2}{2}], \tag{12}$$

$$I(\xi_1, t) = J(\xi_1, t) \exp\,[-\frac{\beta b''(t)}{2D_{\xi_2}^3} [g(0) - g(\xi_2)] \frac{\xi_1^3}{6}]. \tag{13}$$

Assuming arbitrary functions of variable $t - \xi_1/D_{\xi_2}$, the probability density $J^{st}(\xi_1, t)$ becomes $\Theta[t - \xi_1/D_{\xi_2}]$, and the higher-order terms proportional to $1/\tau^3$ are neglected. Therefore, we have

$$F(\xi_1, t) = \Theta[t - \xi_1/D_{\xi_2}] I^{st}(\xi_1, t) H^{st}(\xi_1, t) F^{st}(\xi_1, t). \tag{14}$$

By a similar method, from Eq. (10) to Eq. (14), we also obtain the Fourier transform of the probability density for $\xi_2$ as

$$F(\xi_2,t) = \Theta[t-\xi_2/D_{\xi_1}]I^{st}(\xi_2,t)H^{st}(\xi_2,t)F^{st}(\xi_2,t). \tag{15}$$

Calculating Eqs. (14) and (15), the Fourier transform of the joint probability density is

$$F(\xi_1,\xi_2,t) = F(\xi_1,t)F(\xi_2,t) = exp[-\frac{\beta t^3}{4[2\pi]^{1/2}\sigma_0^3\tau}\xi_2^2]. \tag{16}$$

By taking the inverse Fourier transform, we obtain

$$F\left(p_{x_2},t\right) = [[2\pi]^{1/2}\frac{\beta t^3}{2\sigma_0^3\tau}]^{-1/2}\, exp[-\frac{2\pi]^{1/2}\sigma_0^3\tau}{\beta t^3}p_{x_2}^2]. \tag{17}$$

The mean square momentum for $F(p_{x_2},t)$ are

$$< p_{x_2}^2(t) >= \frac{\beta}{2[2\pi]^{1/2}\sigma_0^3\tau}t^3. \tag{18}$$

In the long-time regime, we find the probability densities $F(p_{x_2},t)$and $F(p_{x_2},t)$ in the long-time regime. We write an approximate equation from Eqs. (8) and (9) as

$$\frac{\partial}{\partial t}F(\xi_1,t) \cong \frac{\beta}{2}b(t)[g(\xi_2)-g(0)]F(\xi_1,t), \tag{19}$$

$$\frac{\partial}{\partial t}F(\xi_2,t) \cong \frac{\beta}{2}b(t)[g(\xi_2)-g(0)]F(\xi_2,t), \tag{20}$$

so that

$$F_{\xi_1}(\xi_1,t) \cong \exp\,(\frac{\beta}{2}\int b(t)[g(\xi_2)-g(0)]\,dt). \tag{21}$$

Using $\int b(t)dt = t-\tau$ (with $b(t)=1$) and setting $F_{\xi_1}(\xi_1,t) = H_{\xi_1}(\xi_1,t)F_{\xi_1}^{st}(\xi_1,t)$, we obtain

$$H_{\xi_1}^{st}(\xi_1,t) \cong \exp\,(-\frac{\beta}{2}\int b(t)[g(\xi_2)-g(0)]\,dt). \tag{22}$$

Hence, in the long-time regime,

$$F(\xi_1,t) = \Theta[t-\xi_1/D_{\xi_2}]\, H_{\xi_1}^{st}(\xi_1,t)F^{st}(\xi_1,t). \tag{23}$$

Using a similar procedure from Eqs. (33) and (35) for $\xi_2$, $F(\xi_2,t)$ is derived as

$$F(\xi_2,t) = \Theta[t-\xi_2/D_{\xi_1}]\, H_{\xi_2}^{st}(\xi_2,t)F^{st}(\xi_2,t). \tag{24}$$

Therefore, combining Eqs. (23) and (24), we calculate that

$$F(\xi_1,\xi_2,t) = F(\xi_1,t)F(\xi_2,t) = exp[-\frac{\beta t^2}{4[2\pi]^{1/2}\sigma_0^3\tau}\xi_2^2]. \tag{25}$$

Using the inverse Fourier transform, the long-time density is

$$F(p_{x_2},t) = [[2\pi]^{1/2}\frac{\beta t^2}{2\sigma_0^3\tau}]^{-1/2}\, exp[-\frac{[2\pi]^{1/2}\sigma_0^3\tau}{\beta t^2}p_{x_2}^2] \tag{26}$$

with the mean square momentum $< p_{x_2}^2(t) >= \frac{\beta}{2[2\pi]^{1/2}\sigma_0^3\tau}t^2$.

For $\tau = 0$, the time derivatives of probability densities $F(\xi_1,t)$ and $F(\xi_2,t)$ from Eqs. (8) and (9) for $\xi_1$ and $\xi_2$ reduce to

$$\frac{\partial}{\partial t}F(\xi_1,t) = -\frac{D_{\xi_2}}{2}\frac{\partial}{\partial \xi_1}F(\xi_1,t) + \frac{\beta}{2}b(t)[g(\xi_2)-g(0)]F(\xi_1,t), \tag{27}$$

$$\frac{\partial}{\partial t}F(\xi_2,t) = -\frac{D_{\xi_2}}{2}\frac{\partial}{\partial \xi_2}F(\xi_1,t) + \frac{\beta}{2}b(t)[g(\xi_2)-g(0)]F(\xi_2,t), \tag{28}$$

where $b(t)=1$. In the steady state, we calculate $F^{st}(\xi_1,t)$ and $F^{st}(\xi_2,t)$ as

$$F^{st}(\xi_1,t) = \exp\,[\frac{\beta}{2D_{\xi_2}}[g(\xi_2)-g(0)]\xi_1],\;\; F^{st}(\xi_2,t) \cong \exp\,[\frac{\beta}{2D_{\xi_1}}\int[g(\xi_2)-g(0)]\,d\xi_2. \tag{29}$$

From Eq. (29), dependent forms are

$$F(\xi_1, t) = \Theta[t - \xi_1/D_{\xi_2}]F^{st}(\xi_1, t),\ \ F(\xi_2, t) = \Theta[t - \xi_2/D_{\xi_1}]F^{st}(\xi_2, t). \tag{30}$$

Therefore, combining $F(\xi_1, t)$ and $F(\xi_2, t)$ from Eq. (30), we calculate that

$$F(\xi_1, \xi_2, t) = F(\xi_1, t)F(\xi_2, t) = exp[-\frac{\beta t^2}{2[2\pi]^{1/2}\sigma_0^3\tau}\xi_2^2]. \tag{31}$$

Using the inverse Fourier transform, the long-time density is

$$F(p_{x_2}, t) = [[2\pi]^{1/2}\frac{\beta t^2}{\sigma_0^3\tau}]^{-1/2}\, exp[-\frac{[2\pi]^{1/2}\sigma_0^3\tau}{2\beta t^2}p_{x_2}^2]. \tag{32}$$

Consequently, the mean square momentum for $F(p_{x_2}, t)$ is obtained as

$$< p_{x_2}^2(t) >= \frac{\beta}{[2\pi]^{1/2}\sigma_0^3\tau}t^2. \tag{33}$$

The mean square momentum scales as $\sim t^2$ for a quantum particle with a correlated Gaussian noise in $t \ll \tau$ and as $\sim t$ in $t \to \infty$, respectively. In the short-time regime, the mean square momentum is proportional to $t^2$ and $t^3$ in the case of independent derivatives of $\xi_1$ and the case of including the $\xi_1 - \xi_2$ mixed derivatives, respectively. Consequently, an initial Gaussian state remains Gaussian at all times, Because the evolution operator is quadratic in the conjugate variables.

The Schrödinger equation for a time-dependent particle in position space is:

$$i\hbar\frac{\partial}{\partial t}\Psi(x, t) = -\frac{\hbar^2}{2m}\frac{\partial^2}{\partial x^2}\Psi(x, t) + V(x, t)\Psi(x, t). \tag{34}$$

The Schrödinger equation itself is a differential equation that describes how a wave function evolves over time. We assume to be correlated that a quantum particle moves in a corelated Gaussian force $V(x, t)$ in one space dimension with vanishing mean value and space-time correlation

$$< V(x, t)V(x', t') >= \frac{V_0^2}{2\tau}f(x - x')\, exp(-|\, t - t'|/\tau). \tag{35}$$

We have the pure state density matrix $\rho(x', x, t{:}V) = \Psi^*(x', t)\Psi(x, t)$ and $< \rho(x', x, t) >_V$ is dependent of the noise realization [13]. Thus, we get the equation of motion with Eq. (35) as

$$\frac{\partial}{\partial t} < \rho\left(x', x, t\right) >_V = -\frac{i\hbar}{2m}\left(\frac{\partial^2}{\partial x^2} - \frac{\partial^2}{\partial x'^2}\right) < \rho\left(x', x, t\right) >_V + \alpha a(t)f(x - x') < \rho\left(x', x, t\right) >_V, \tag{36}$$

where $\alpha = \frac{V_0^2}{2\hbar^2}$ and the parameter denote $a(t) = 1 - \exp{(-t/\tau)}$ for a quantum particle with a correlated Gaussian noise.

For the short-time regime $t \ll \tau$ and $a(t) = t/\tau$ , we find the mean square values from $< \rho(x', x, t) >_V$. The initial condition of a particle is assumed to be in a wave packet centered at the origin $x = 0$ We put $< \rho(x', x, 0) >_V = \Psi^*(x', 0)\Psi(x, 0)$, having $\Psi(x, 0) = [(2\pi)^{\frac{1}{4}}\sigma_1^{\frac{1}{2}}]^{-1}\exp\left(-\frac{x^2}{4\sigma_1^2}\right)$.

Introducing the characteristic coordinates $x_1 = x + x'$, $x_2 = x - x'$ and taking the Laplace transform in Eq. (36), we get the reduced equations of motion as

$$\frac{2i\hbar}{m}\frac{\partial^2}{\partial x_1\partial x_2}\overline{\rho}(x_1, x_2, s) = -\left(s - \frac{\alpha}{s^2}f(x_2)\right)\overline{\rho}(x_1, x_2, s) + \overline{\rho}(x_1, x_2, 0), \tag{37}$$

where $< \rho(x', x, t) >_V \equiv \overline{\rho}(x', x, t)$. We define the Fourier transform and the Laplace transform as

$$\overline{\rho}(k_1, k_2, t) = \int_{-\infty}^{+\infty} dx\overline{\rho}(x_1, x_2, t)\exp{(-ik_1x_1 - ik_2x_2)} \tag{38}$$

and

$$\overline{\rho}(x_1, x_2, s) = \int_{-\infty}^{+\infty} dt\overline{\rho}(x_1, x_2, t)\exp{(-st)}. \tag{39}$$

Taking the Fourier transform on both sides of the above equation and doing some calculations, we obtain the following $\overline{\rho}(k_1, x_2, s)$ as

$$\overline{\rho}(k_1, x_2, s) = \int_0^{+\infty} dzexp[-\left(2\sigma^2 + \frac{\hbar^2 x_2^2}{2m^2\sigma^2}\right)k_1^2 - sz]\exp{[-\frac{\alpha}{\tau s^2}\int_0^{x_2} dz' f(cz')]}. \tag{40}$$

Here, $c = 2\hbar k/m$ Since the right-hand side of this equation is in the form of a Laplace transform, we take on inversion

$$\overline{\rho}(k_1, 0, t) = 2exp\left[-\left(2\sigma^2 + \frac{\hbar^2 t^2}{2m^2\sigma^2}\right)k_1^2\right]\exp\left[\frac{\alpha t}{\tau}\int_0^t dz' f(cz')\right], \tag{41}$$

where the function denotes $f(z) = [2\pi\sigma_1^2]^{-1/2}\exp(-z^2/2\sigma_1^2)$ to be Gaussian, $\sigma_0^2 = m^2/8\hbar^2$. The mean square displacement can be expressed as

$$< x_1^2(t) > = -[\frac{\partial^2}{\partial k_1^2}\overline{\rho}(k_1, 0, t)]_{k_1=0}. \tag{42}$$

We have the mean square displacements for a correlated Gaussian noise [14] in the three-time ($t \ll \tau$, $t \gg \tau$, and for $\tau = 0$) regimes as follows: For the short-time regime $t \ll \tau$ with $a(t) = t/\tau$ , the mean square displacement is obtained as

$$< x_1^2(t) > = \sigma_1^2 + \frac{\hbar^2}{4m^2\sigma_1^2}t^2 + \frac{32\hbar^4\alpha}{3\sigma_2^2\tau}t^4. \tag{43}$$

In $t \gg \tau$, we obtain the mean square displacement for a quantum particle with a random noise as

$$< x_1^2(t) > = \sigma_1^2 + \frac{\hbar^2}{4m^2\sigma_1^2}t^2 + \frac{32\hbar^4\alpha}{3\sigma_2^2}t^3. \tag{44}$$

For $\tau = 0$ ($t \to \infty$), the mean square displacement is the same value as Eq. (44).

Consequently, the mean square displacement scales as $\sim t^3$ and $\sim t$ for a quantum particle with a correlated Gaussian noise in $t \ll \tau$ and $\tau = 0$ ($t \to \infty$), respectively. For a random noise, the mean square displacement is proportional to $t^3$ for a random noise when $t \to \infty$.

## 3. Statistical quantities

We calculate the non-Gaussian parameter, the correlation coefficient, the entropy, and the combined entropy for the displacement and the velocity. First of all, the non-Gaussian parameter for displacement and velocity are, respectively, given by

$$K_x = < x^4 > /3 < x^2 >^2, K_{p_x} = < p_x^4 > /3 < p_x^2 >^2. \tag{45}$$

We introduce the correlation coefficient as

$$\rho_{x,p_x} = x_0 p_{x0}/\sigma_x \sigma_{p_x}. \tag{46}$$

Correlation coefficient $\boldsymbol{\rho_{x,p_x}}$ means the statistical quantity describing the strength and direction of a relationship between two variables $x$ and $p_x$. Here, we assume that a quantum particle is initially at $x = x_0$ and at $p_x = p_{x0}$. $\sigma_x$ and $\sigma_{p_x}$ denote the root-mean-squared displacement and the root-mean-squared velocity of the joint probability density, respectively. The entropies are, respectively, calculated as

$$S(x,t) = -\rho(x,t)ln\rho(x,t),\ S(p_x,t) = -F(p_x,t)lnF(p_x,t), \tag{47}$$

and the combined entropy $S(x, p_x, t)$ is defined by

$$S(x, p_x, t) = -\rho(x,t)F(p_x,t)ln\rho(x,t)F(p_x,t). \tag{48}$$

**Table 1** Values of the non-Gaussian parameter, the correlation coefficient, the entropy, and the combined entropy for the motion of a quantum particle with a correlated Gaussian noise in the three-time regimes. Here, we assume that a quantum particle is initially at $x = x_0$ and at $p_x = p_{x0}$.

| Time | $x,\ p_x$ | $K_x, K_{p_x}$ | $\rho_{x,p_x}$ | $S(x,t)$, $S(p_x,t)$ | $S(x,p_x,t)$ |
|---|---|---|---|---|---|
| $t \ll \tau$ | $x$ | $\frac{\sigma_2^4\tau^2 x_0^4}{\hbar^8\alpha_1^2}t^{-8} + \frac{\sigma_2^2\tau x_0^2}{\hbar^4\alpha}t^{-4}$ | $\frac{\sigma_2\sigma_0^{3/2}\tau x_0 p_{x0}}{\hbar^2\alpha^{1/2}\beta^{1/2}}t^{-7/2}$ | $ln\frac{\hbar^4\alpha}{\sigma_2^2\tau}t^4$ | $ln\frac{\hbar^4\alpha\beta}{\sigma_0^3\sigma_2^2\tau^2}t^7$ |
| | $p_x$ | $\frac{\sigma_0^6\tau^2 p_{x0}^4}{\beta^2}t^{-6} + \frac{\sigma_0^3\tau p_{x0}^2}{\beta}t^{-3}$ | | $ln\frac{\beta}{\sigma_0^3\tau}t^3$ | |
| $t \gg \tau$ | $x$ | $\frac{\sigma_2^4 x_0^4}{\hbar^8\alpha_1^2}t^{-6} + \frac{\sigma_2^2 x_0^2}{\hbar^4\alpha}t^{-3}$ | $\frac{\sigma_2\sigma_0^{3/2}\tau^{1/2} x_0 p_{x0}}{\hbar^2\alpha^{1/2}\beta^{1/2}}t^{-5/2}$ | $ln\frac{\hbar^4\alpha}{\sigma_2^2}t^3$ | $ln\frac{\hbar^4\alpha\beta}{\sigma_0^3\sigma_2^2\tau}t^5$ |
| | $p_x$ | $\frac{\sigma_0^6\tau^2 p_{x0}^4}{\beta^2}t^{-4} + \frac{\sigma_0^3\tau p_{x0}^2}{\beta}t^{-2}$ | | $ln\frac{\beta}{\sigma_0^3\tau}t^2$ | |
| $\tau = 0$ | $x$ | $\frac{\sigma_2^4 x_0^4}{\hbar^8\alpha_1^2}t^{-6} + \frac{\sigma_2^2 x_0^2}{\hbar^4\alpha}t^{-3}$ | $\frac{\sigma_2\sigma_0^{3/2}\tau^{1/2} x_0 p_{x0}}{\hbar^2\alpha^{1/2}\beta^{1/2}}t^{-5/2}$ | $ln\frac{\hbar^4\alpha}{\sigma_2^2}t^3$ | $ln\frac{\hbar^4\alpha\beta}{\sigma_0^3\sigma_2^2\tau}t^5$ |
| | $p_x$ | $\frac{\sigma_0^6\tau^2 p_{x0}^4}{\beta^2}t^{-4} + \frac{\sigma_0^3\tau p_{x0}^2}{\beta}t^{-2}$ | | $ln\frac{\beta}{\sigma_0^3\tau}t^2$ | |

## 4. Summary

In summary, we have investigated the diffusive behaviors of a quantum particle driven by correlated Gaussian noises [14,15]. We obtained the analytical solution of the joint probability density function and derived the corresponding mean squared values. In particular, the analytical expressions were explicitly derived in the three-time regimes $t \ll \tau$, $t \gg \tau$ and $\tau = 0$ $(t \to \infty)$, where $\tau$ denotes the correlation time of the noise.

The main findings are as follows: (i) The mean square momentum exhibits a super-diffusive scaling proportional to $t^3$ in the short-time regime, while it crosses over to a normal diffusive behavior proportional to $t^2$ in the long-time limit. In the short-time regime, the scaling behavior depends crucially on whether the mixed derivative term $\partial_{\xi_1} \partial_{\xi_2}$ is included. Without the mixed derivative, the mean square momentum scales as $t^2$, whereas the inclusion of the mixed derivative leads to the enhanced $t^3$ scaling. (ii) For the short-time regime $t \ll \tau$, the mean square displacement is obtained as $t^4$, and the mean square displacement in $\tau = 0$ $(t \to \infty)$ is the same value as the result of Ref. [4]. (iii) Table 1 is summarized the values of the non-Gaussian parameter, the correlation coefficient, the entropy, and the combined entropy for the motion of a quantum particle in the three-time regimes.

In the case of the field-off initial conditions, a mean square displacement grows as $t^4$ leading in the short-time to giant enhancement of the particle transport [16], consistent with our result. These results clarify the fundamental role of correlated Gaussian noises and mixed derivative structures in determining quantum diffusion characteristics. The present analytical framework may be extended to more general non-Markovian environments [17-19], non-equilibrium quantum transport [20-22], and entropy production problems in open quantum systems [23]. Future work may explore memory-kernel generalizations, fractional correlations, and thermodynamic interpretations of the anomalous short-time super-diffusion identified in this study.

## References

1. J. Heinrichs, Quantum treatment of Brownian motion and influence of dissipation on diffusion in dynamically disordered systems, Z. Phys. B 50, 269 (1983).

2. J. Heinrichs, Quantum transport in dynamically disordered continuum tight binding systems, Z. Phys. B 52, 9 (1983).

3. J. Heinrichs, Continuum tight binding models and quantum transport in the presence of dynamical disorder, Z. Phys. B 53, 175 (1983).

4. A.M. Jayannavar, N. Kumar, Nondiffusive Quantum Transport in a Dynamically Disordered Medium, Phys. Rev. Lett. 48, 553 (1982).

5. A.A. Ovchinnikov, N.S. Erikhman, Equation for the density of states in a one-dimensional random potential, Zh. Eksp. Teor. Fiz. 67, 1474 (1974) [Soy. Phys. JETP 40, 733 (1974)].

6. A. Madhukar, W. Post, Exact Solution for the Diffusion of a Particle in a Medium with Site Diagonal and Off-Diagonal Dynamic Disorder, Phys. Rev. Lett. 39, 1424 (1977).

7. J.-H. Jeon, N. Leijnse, L.B. Oddershede, R. Metzler, Anomalous diffusion and power-law relaxation in wormlike micellar solution, New J. Phys. 15, 045011 (2013).

8. S. Joo, J.H. Jeon, Viscoelastic active diffusion governed by nonequilibrium fractional Langevin equations: Underdamped dynamics and ergodicity breaking. Chaos Solitons Fractals 177, 114288 (2013).

9. C. Lim, J.-H. Jeon, Anomalous diffusion in coupled viscoelastic media: A fractional Langevin equation approach, Phys. Rev. Research 7, 043356 (2025)

10. J.W. Jung, S.K. Seo, K. Kim, Joint probability densities of an active particle coupled to two heat reservoirs. Physica. A 668, 130483 (2025).

11. J.W. Jung, S.K. Seo, K. Kim, Dynamical behavior of passive particles with harmonic, viscous, and correlated Gaussian forces, Phys. Lett. A 2025, 546, 130512 (2025).

12. Y.J. Kang, J.W. Jung, S.K. Seo, K. Kim, On the Stochastic Motion Induced by Magnetic Fields in Random Environments. Entropy 27, 330 (2025).

13. E.A. Novikov, Functionals and the Random-Force Method in Turbulence Theory, Zh. Eksp. Teor. Fiz. 47,

1919 (1964) [Soy. Phys. JETP 20, 1290 (1965)].

14. J. Heinrichs, Diffusion and superdiffusion of a quantum particle in time-dependent random potentials, Z. Phys. B 89, 115-121 (1992).

15. N. Kumar, in Stochastic Processes: Formalism and Applications (Lecture Notes in Physics, Springer: Berlin, Heidelberg, New York, 1983).

16. D. Gamba, B. Cui, A. Zaccone, Hyperballistic transport in dense systems of charged particles under AC electric fields, arXiv:2402.08519v3.

17. J. Zou, S. Bosco, D. Loss, Spatially correlated classical and quantum noise in driven qubits, npj Quantum Information 10, 46 (2024).

18. A. Kurt, Interplay between Non-Markovianity of Noise and Quantum Dynamics, Entropy 25, 501 (2023).

19. A. Amir, Y. Lahini, H.B. Perets, Classical diffusion of a quantum particle in a noisy environment, Phys. Rev. E 79, 050105R (2009).

20. K. Luoma, W.T. Strunz, J. Piilo, Diffusive Limit of Non-Markovian Quantum Jumps, Phys. Rev. Lett. 125, 150403 (2020).

21. D.S. Bhakuni, T.L.M. Lezama, Y.B. Lev, Noise-induced transport in the Aubry-André-Harper model, SciPost Phys. Core 7, 023 (2024).

22. X. Zhang, Z. Wu, G.A.L. White, et al. Learning and forecasting open quantum dynamics with correlated noise, *Communications Physics* 8, 29 (2025).

23. D. Gamba, B. Cui, A. Zaccone, Open quantum systems with particle and bath driven by time-dependent fields, arXiv:2505.05348v2.